\documentstyle[12pt]{article}
\newskip\humongous \humongous=0pt plus 1000pt minus 1000pt

\newif\ifdtup

\def\abs#1{\left| #1\right|}
\def\pr#1{#1^\prime}

\def\beq{\begin{equation}}
\def\eeq{\end{equation}}

\def\beqn{\begin{eqnarray}}
\def\eeqn{\end{eqnarray}}
\relax

\def\dotx{\dotx{\dot\overline{x}}}

\relax

\catcode`\@=11

\@addtoreset{equation}{section}
\def\theequation{\thesection\arabic{equation}}

\def\@normalsize{\@setsize\normalsize{15pt}\xiipt\@xiipt
\abovedisplayskip 14pt plus3pt minus3pt%
\belowdisplayskip \abovedisplayskip
\abovedisplayshortskip \z@ plus3pt%
\belowdisplayshortskip 7pt plus3.5pt minus0pt}

\def\small{\@setsize\small{13.6pt}\xipt\@xipt
\abovedisplayskip 13pt plus3pt minus3pt%
\belowdisplayskip \abovedisplayskip
\abovedisplayshortskip \z@ plus3pt%
\belowdisplayshortskip 7pt plus3.5pt minus0pt
\def\@listi{\parsep 4.5pt plus 2pt minus 1pt
     \itemsep \parsep
     \topsep 9pt plus 3pt minus 3pt}}

\@twosidetrue

\relax

\catcode`@=12

\catcode`\@=11

\def\section{\@startsection{section}{1}{\z@}{3.5ex plus 1ex minus
   .2ex}{2.3ex plus .2ex}{\large\bf}}

\def\thesection{\arabic{section}.}

\def\appendix{\setcounter{section}{0}
 \def\thesection{APPENDIX \Alph{section}:}
 \def\theequation{\Alph{section}.\arabic{equation}}}

\def\ps@headings{\def\@oddfoot{}\def\@evenfoot{}
\def\@oddhead{\hbox{}\hfill
 \makebox[.5\textwidth]{\raggedright\ignorespaces --\thepage{}--
 \hfill {}}}  
\def\@evenhead{\@oddhead}
\def\subsectionmark##1{\markboth{##1}{}}
}

\ps@headings

\catcode`\@=12

\def\figcap{\section*{Figure Captions\markboth
 {FIGURECAPTIONS}{FIGURECAPTIONS}}\list
 {Fig. \arabic{enumi}:\hfill}{\settowidth\labelwidth{Fig. 999:}
 \leftmargin\labelwidth
 \advance\leftmargin\labelsep\usecounter{enumi}}}
 \relax
\def\tablecap{\section*{Table Captions\markboth
 {TABLECAPTIONS}{TABLECAPTIONS}}\list
 {Table \arabic{enumi}:\hfill}{\settowidth\labelwidth{Table 999:}
 \leftmargin\labelwidth
 \advance\leftmargin\labelsep\usecounter{enumi}}}
 \relax
\def\reflist{\section*{References\markboth
 {REFLIST}{REFLIST}}\list
 {[\arabic{enumi}]\hfill}{\settowidth\labelwidth{[999]}
 \leftmargin\labelwidth
 \advance\leftmargin\labelsep\usecounter{enumi}}}
 \relax

\catcode`\@=11

\def\ps@headings{\def\@oddfoot{}\def\@evenfoot{}
\def\@oddhead{\hbox{}\hfill
 \makebox[.5\textwidth]{\raggedright\ignorespaces --\thepage{}--
 \hfill {}}}    
\def\@evenhead{\@oddhead}
\def\subsectionmark##1{\markboth{##1}{}}
}

\ps@headings

\relax

\relax

\def\zp#1#2#3{{\it Z. Phys. }{\bf #1}(19#2)#3}

\def\prep#1#2#3{{\it Phys. Rep. }{\bf #1}(19#2)#3}
\def\pr#1#2#3{{\it Phys. Rev. }{\bf #1}(19#2)#3}
\def\np#1#2#3{{\it Nucl. Phys. }{\bf #1}(19#2)#3}

\relax

\def\theequation{\arabic{equation}}
\newcommand\sss{\scriptscriptstyle}

\newcommand\aem{\alpha_{\rm em}}
\newcommand\WW{Weizs\"acker-Williams}

\begin{document}

\begin{titlepage}
\nopagebreak
\vspace*{-1in}
{\leftskip 11cm
\normalsize
\noindent
\newline
CERN-TH.7032/93\\
GeF-TH-18/93

}
\vfill
\begin{center}
{\large \bf \sc Improving the Weizs\"acker-Williams approximation
in electron-proton collisions}
\vfill
{\bf Stefano Frixione}
\vskip .3cm
{Dip. di Fisica, Universit\`a di Genova, and INFN, Sezione di Genova,
Genoa, Italy}\\
\vskip .6cm
{\bf Michelangelo L. Mangano}
\vskip .3cm
{INFN, Scuola Normale Superiore and Dipartimento di Fisica, Pisa, Italy}\\
\vskip .6cm
{\bf Paolo Nason\footnotemark}
\footnotetext{On leave of absence from INFN, Sezione di Milano, Milan, Italy.}
and
{\bf Giovanni Ridolfi\footnotemark}
\footnotetext{On leave of absence from INFN, Sezione di Genova, Genoa, Italy.}
\vskip .3cm
{CERN TH-Division, CH-1211 Geneva 23, Switzerland}
\end{center}
\vfill
\nopagebreak
\begin{abstract}
{\small
We critically examine the validity of the Weizs\"acker-Williams approximation
in electron-hadron collisions. We show that in its commonly used form it can
lead to large errors, and we show how to improve it in order to get accurate
results. In particular, we present an improved form that is valid beyond the
leading logarithmic approximation in the case when a small-angle cut is applied
to the scattered electron. Furthermore we include comparisons of the
approximate
expressions with the exact electroproduction calculation in the case of
heavy-quark production.
}
\end{abstract}
\vfill
CERN-TH.7032/93
\newline
October 1993    \hfill
\end{titlepage}
Electroproduction phenomena are usually computed by assuming that the incoming
electron beam can be considered to be equivalent to a photon broad-band
beam [\ref{WWpaper}].
The photon distribution in the electron is given by the formula
\beq
\label{WWequation}
f_{\gamma}^{(e)}(y)=\frac{\aem}{2\pi}
\left[ \frac{1+(1-y)^2}{y}\log\frac{(Sy-S_{min})(1-y)}{m^2_ey^2}
+{\cal O}(1) \right],
\eeq
where $S$ is the electron-hadron centre-of-mass energy squared, $S_{min}$ is
the minimum value for the invariant mass squared of the produced hadronic
system, $m_e$ is the electron mass, and $y$ is the fraction of the electron
energy carried by the photon. Several examples of applications of this formula,
as well as discussions of its range of validity, can be found in
ref.~[\ref{HeraWorkshop}]. We have explicitly indicated in
eq.~(\ref{WWequation}) that the neglected terms are of order 1 with respect to
the logarithmically-enhanced one. The logarithmic term is typically of order
15 to 20 at HERA energies, and therefore one expects the error associated with
this approximation to be of the order of 5 to 10\%. It was noted in
ref.~[\ref{CataniHautmann}] that the error can in fact be much larger than
this, unless one takes appropriately into account the
dynamical effects that modify
the ``effective'' upper scale entering the logarithm in
eq.~(\ref{WWequation}).

In this paper, we show that several modifications to
eq.~(\ref{WWequation}) are needed in order to meet the needs of the
photoproduction experimental conditions one encounters at HERA. There a
photoproduction event is usually defined with an appropriate anti-tag
condition (i.e. outgoing electrons above a given angular cut are
vetoed). Therefore, we reconsider the derivation of
the Weizs\"acker-Williams distribution, taking into account the particular
experimental conditions of HERA. First of all, we derive a \WW\ formula for the
case when a small-angle cut is applied to the outgoing electron, which is valid
beyond the leading logarithmic approximation.

Let us consider the electroproduction process
\beq
e(p)+p(k)\to e(p^\prime)+X,
\eeq
where $p$ is a massless parton ($k^2=0$), $X$ is a generic hadronic system,
and $p^2 = {p^\prime}^2 = m_e^2$.
The cross section for the process is given by
\beq
\label{sigmaep}
d\sigma_{ep}=\frac{1}{8k\cdot p}\frac{e^2W^{\mu\nu}T_{\mu\nu}}{q^4}
\frac{d^3 p^\prime}{(2\pi)^3 2E^\prime}\,\,,
\eeq
where
\beq
q=p-p^\prime.
\eeq
The tensor $T_{\mu\nu}$ is the electron tensor, given by
\beq\label{tmunu}
T_{\mu\nu}=4\left(\frac{1}{2}q^2g_{\mu\nu}
+p_\mu p^\prime_\nu+p_\nu p^\prime_\mu\right).
\eeq
Notice that
\beq
q^\mu T_{\mu\nu} = q^\nu T_{\mu\nu} = 0,
\eeq
since $p\cdot q = -p^\prime \cdot q = q^2/2$.

The tensor $W^{\mu\nu}$ is the hadron tensor. Exploiting
current conservation, which requires $q_\mu W^{\mu\nu} = q_\nu W^{\mu\nu} = 0$,
it can be decomposed as
\beq
\label{wmunu}
W^{\mu\nu}=
W_1(q^2,k\cdot q)\left(-g^{\mu\nu}+\frac{q^\mu q^\nu}{q^2}\right)-
\frac{q^2}{(k\cdot q)^2}W_2(q^2,k\cdot q)
\left(k^\mu-\frac{k\cdot q}{q^2}q^\mu\right)
                 \left(k^\nu-\frac{k\cdot q}{q^2}q^\nu\right)
\eeq
(a term proportional to $\epsilon_{\mu\nu\rho\sigma}q^\rho k^\sigma$
has not been included, since it gives zero contribution in this case).
In the limit $q^2\to 0$, $W^{\mu\nu}$ must be an analytic function of $q^2$. By
requiring that $q^2 W^{\mu\nu}$ vanish for $q^2=0$, we obtain
\beq
\label{w1w2}
W_2(q^2,k\cdot q)=W_1(0,k\cdot q)+{\cal O}(q^2).
\eeq

We drop for the moment the ${\cal O}(q^2)$ terms. We will discuss later their
effect, which in general leads to non-universal and non-logarithmic corrections
to the factorised Weizs\"acker-Williams approximation.
{}From eqs.~(\ref{tmunu}), (\ref{wmunu}) and (\ref{w1w2}) we find
\beq
\label{wmunutmunu}
W^{\mu\nu}T_{\mu\nu}=
-4 W_1(0,k\cdot q)\left[2m_e^2+q^2\frac{1+(1-y)^2}{y^2}\right],
\eeq
where
\beq
y=\frac{k\cdot q}{k\cdot p}=1-\frac{k\cdot p^\prime}{k\cdot p}.
\eeq
Notice that $y$ is precisely
the fraction $z$ of longitudinal momentum carried by the photon. In fact,
if $k=k^0(1,0,0,-1)$ and $q=zp+\hat{q}$, with $k\cdot \hat{q}=0$
one finds $y=(k\cdot q)/(k\cdot p)=z$.

We must now find a convenient expression for the phase space of the scattered
electron. We have
\beq
\frac{d^3 p^\prime}{E^\prime}=2\pi{\beta^\prime}^2 E^\prime dp^\prime
d\cos\theta,
\eeq
where the trivial azimuth integration has already been carried out, and
\beqn
&&p^\prime=(E^\prime,0,E^\prime\beta^\prime\sin\theta,
E^\prime\beta^\prime\cos\theta)
\\
&&p=(E,0,0,E\beta)
\\
&&\beta^\prime=\sqrt{1-\frac{m_e^2}{{E^\prime}^2}}, \;\;\;
\beta=\sqrt{1-\frac{m_e^2}{{E}^2}}.
\eeqn
We now trade the two variables $(p^\prime,\cos\theta)$ for
$(q^2,y)$. From the definitions given above, we find
\beqn
\label{qsquare}
&&q^2=2m_e^2-2EE^\prime
\left(1-\beta\beta^\prime\cos\theta\right)
\\
\label{y}
&&y=1-\frac{E^\prime\left(1+\beta^\prime\cos\theta\right)}{E(1+\beta)},
\eeqn
and it is easy to prove that the Jacobian of this change of variables
is simply $2E^\prime{\beta^\prime}^2$. Therefore
\beq
\frac{d^3 p^\prime}{E^\prime}=\pi dq^2 dy.
\eeq

We are now ready to compute the cross section in eq.~(\ref{sigmaep}). We get
\beq
d\sigma_{ep}=
-\frac{\aem}{2\pi}\frac{W_1(0,k\cdot q)}{4k\cdot p}
\left[\frac{2m_e^2}{q^4}+\frac{1+(1-y)^2}{y^2q^2}\right]dq^2 dy,
\eeq
where, as usual, $\aem=e^2/(4\pi)$. Integration in $q^2$ gives
\beq
d\sigma_{ep}=
\sigma_{\gamma p}(q,k)f_\gamma^{(e)}(y) dy,
\eeq
where
\beq
f_\gamma^{(e)}(y)=
\frac{\aem}{2\pi}\left[
2m_e^2y\left(\frac{1}{q^2_{max}}-\frac{1}{q^2_{min}}\right)
+\frac{1+(1-y)^2}{y}
\log\frac{q^2_{min}}{q^2_{max}}\right],
\eeq
and
\beq
\label{sigmagammap}
\sigma_{\gamma p}(q,k)=
-\frac{g_{\mu\nu}W^{\mu\nu}}{8k\cdot q}
=\frac{W_1(0,k\cdot q)}{4k\cdot q}
\eeq
is the cross section
for the process $\gamma(q)+p(k) \to X$
for an on-shell photon.

We must determine the two integration bounds $q^2_{max}$ and $q^2_{min}$.
For $\theta\ll 1$ we have, from eq.~(\ref{y}),
\beq
\label{oldy}
E^\prime = \frac{A^2+m_e^2}{2A} + \frac{(A^2-m_e^2)^2}{8A^3}\theta^2
+ {\cal O}(\theta^4),
\eeq
where
\beq
A=E(1+\beta)(1-y),
\eeq
and eq.~(\ref{qsquare}) becomes
\beq\label{qsquareapp}
q^2=-\frac{m_e^2 y^2}{1-y}-\frac{E(1+\beta)(A^2-m_e^2)^2}{4A^3}\theta^2
+{\cal O}(\theta^4).
\eeq
The value of $q^2_{max}$ is obtained by taking $\theta=0$, namely
\beq
q^2_{max}=-\frac{m_e^2 y^2}{1-y}.
\eeq
Analogously, the value of $q^2_{min}$ is obtained when $\theta$ equals its
maximum value $\theta_c$. If $\theta_c \ll 1$ we can use eq.~(\ref{qsquareapp})
to obtain
\beqn
\label{virtmin}
q^2_{min}
&=&-\frac{m_e^2 y^2}{1-y}-\frac{E(1+\beta)(A^2-m_e^2)^2}{4A^3}\theta_c^2
+{\cal O}(\theta^4)
\nonumber \\
&=&
-\frac{m_e^2 y^2}{1-y}-E^2(1-y)\theta_c^2
+{\cal O}(E^2\theta_c^4,m_e^2\theta_c^2,m_e^4/E^2).
\eeqn
The function $f_\gamma^{(e)}(y)$ takes the form
\beqn      \label{ourww}
f_\gamma^{(e)}(y)&=&\frac{\aem}{2\pi}\left\{
2(1-y)
\left[\frac{m_e^2y}{E^2(1-y)^2\theta_c^2 + m_e^2y^2}-\frac{1}{y}\right]\right.
\nonumber \\
&&\left.+\frac{1+(1-y)^2}{y}
\log\frac{E^2(1-y)^2\theta_c^2 +m_e^2 y^2}{m_e^2 y^2}
+{\cal O}(\theta_c^2,m_e^2/E^2)\right\}.
\eeqn
Observe that the uncertainty is small with respect to 1. Also notice that the
non-logarithmic term is singular in $y$ and therefore represents a
non-negligible correction. This subleading term is universal, in the sense that
it is independent of the particular hard process we are considering.

We now comment on the effect of the ${\cal O}(q^2)$ terms in
eq.~(\ref{w1w2}).  It is straightforward to verify that these terms will add a
contribution to the differential cross section given by
\beq
\Delta d\sigma_{ep} = \frac{\aem}{2\pi}\frac{1}{2k\cdot q}
\left[\sum_{n=0} A^{(n)}(k\cdot q)\left(\frac{q^2}{q\cdot k}\right)^n
\frac{dq^2}{q\cdot k}\right]
\frac{dy}{y},
\eeq
where the coefficients $A^{(n)}(q\cdot k)$ are obtained from the derivatives of
the $W_2$ form factor evaluated at $q^2=0$, and are of order 1.
For simplicity we neglected terms that are finite for $y\to 0$, proportional
to derivatives of both $W_1$ and $W_2$,  and for which the following
conclusions hold as well.
Upon integration over $q^2$, we are left with
\beq  \label{sublead}
\Delta d\sigma_{ep} = \frac{\aem}{2\pi}\frac{1}{2k\cdot q}
\left[\sum_{n=1} \frac{A^{(n-1)}}{n}
\left(\frac{q^2_{min}}{q\cdot k}\right)^n\right]
\frac{dy}{y},
\eeq
where we neglected the contributions proportional to $q^2_{max}$, which are
suppressed by powers of $m_e^2/s$.
In the case of a small angular cut we have
\beq
\label{q2ltth2}
\frac{\abs{q^2_{min}}}{2q\cdot k} < \frac{E^2\theta_c^2}{S_{min}}.
\eeq
For example, in the case of charm production and using $E=30$ GeV,
$\theta_c=5\times 10^{-3}$ we would have
$\abs{q^2_{min}}/(2q\cdot k) < 2.5 \times 10^{-3}$.
Therefore  these non-factorisable corrections are negligible, at least
formally, with respect to the factorisable ones.

If no angular cut is applied, however, the ratio $\abs{q^2_{min}}/(2q\cdot k)$,
although limited from above by 1, can be of order 1. In this case,
non-factorisable corrections to the Weizs\"acker-Williams approximation are of
order $1/y$.

As an application of the formulae discussed above, we consider the case of
heavy-quark electroproduction. The total cross section for this process has
recently been calculated at next-to-leading order in QCD by E.~Laenen et al. in
ref.~[\ref{Laenen}]. For simplicity we will perform our comparisons between
exact and approximated results at the leading order in $\alpha_{\sss S}$. For
our goals this is not a limitation, as higher-order QCD corrections should not
affect the properties of the photon density inside the electron and therefore
will not change our results qualitatively.

In table~\ref{uno} we show the total production cross section of charm and
bottom quark pairs at HERA ($E=30$ GeV, $\sqrt{S}=314$ GeV),
with the application of different angular cuts. We
used $m_c=1.5$ GeV, $m_b=4.75$ GeV, and the MRSD0 parton distributions
[\ref{MRSD0}] for the proton. The matrix elements for the exact leading-order
calculation are taken from ref.~[\ref{Laenen}].

As can be seen from table~\ref{uno}, the presence of the
non-logarithmic term improves the agreement between the exact result and the
Weizs\"acker-Williams approximation from the level of 6--7\% to the level of
1\% in the case of charm, and from 5--7\% to 0.1\% the case of the bottom
(apart from the case $\theta_c=0.5$, which is too large for the small-angle
approximation to work properly).

Table~\ref{due} is similar, but evaluated for a configuration with
electron beam energy $E=100$ GeV and $\sqrt{S}$=1 TeV.
We see that with the same angular cuts but higher energy, the approximation
becomes worse (this can easily be explained from eq.~(\ref{q2ltth2})).

\begin{table}
\begin{center}
\begin{tabular}{|l||c|c|c|c|c|} \hline
& \multicolumn{5}{c|}{Charm}
\\ \hline
$\theta_c$
& $\sigma_{\rm exact}$ (nb)
& $\sigma_{\sss WW}$ only log (nb)
& (\%)
& $\sigma_{\sss WW}$ (nb)
& (\%)
\\ \hline \hline
$5\times 10^{-1}$ & 353.3 & 420.7 & 19.08 & 405.5 & 14.77
 \\ \hline
$5\times 10^{-2}$ & 325.4 & 345.2 &  6.08 & 330.0 & 1.41
 \\ \hline
$5\times 10^{-3}$ & 252.9 & 269.8 &  6.68 & 254.6 & 0.67
 \\ \hline
\hline
& \multicolumn{5}{c|}{Bottom}
\\ \hline
$\theta_c$
& $\sigma_{\rm exact}$ (nb)
& $\sigma_{\sss WW}$ only log (nb)
& (\%)
& $\sigma_{\sss WW}$ (nb)
& (\% )
\\ \hline \hline
$5\times 10^{-1}$ & 3.110 & 3.404 & 9.45 & 3.277 & 5.37
 \\ \hline
$5\times 10^{-2}$ & 2.599 & 2.732 & 5.12 & 2.605 & 0.23
 \\ \hline
$5\times 10^{-3}$ & 1.931 & 2.060 & 6.68 & 1.933 & 0.10
 \\ \hline
\end{tabular}
\caption[]{\label{uno}
Total cross sections for charm and bottom production at HERA ($E=30$ GeV,
$\sqrt{S}=314$ GeV) for various angular cuts. The percentual differences
between the exact results and the two approximate ones are also shown.
Statistical errors are smaller that the last digit.}
\end{center}
\end{table}

\begin{table}
\begin{center}
\begin{tabular}{|l||c|c|c|c|c|} \hline
& \multicolumn{5}{c|}{Charm}
\\ \hline
$\theta_c$
& $\sigma_{\rm exact}$ (nb)
& $\sigma_{\sss WW}$ only log (nb)
& (\%)
& $\sigma_{\sss WW}$ (nb)
& (\%)
\\ \hline \hline
$5\times 10^{-1}$ & 660.8 & 866.4 & 31.11 & 839.0 & 26.97
 \\ \hline
$5\times 10^{-2}$ & 650.7 & 732.7 & 12.60 & 705.3 &  8.39
 \\ \hline
$5\times 10^{-3}$ & 549.6 & 599.0 &  8.99 & 571.6 &  4.00
 \\ \hline
\hline
& \multicolumn{5}{c|}{Bottom}
\\ \hline
$\theta_c$
& $\sigma_{\rm exact}$ (nb)
& $\sigma_{\sss WW}$ only log (nb)
& (\%)
& $\sigma_{\sss WW}$ (nb)
& (\%)
\\ \hline \hline
$5\times 10^{-1}$ & 8.585 & 10.20 & 18.81 & 9.866 & 14.92
 \\ \hline
$5\times 10^{-2}$ & 7.997 & 8.492 &  6.19 & 8.154 &  1.96
 \\ \hline
$5\times 10^{-3}$ & 6.360 & 6.780 &  6.60 & 6.441 &  1.27
 \\ \hline
\end{tabular}
\caption[]{\label{due}
Total cross sections for charm and bottom production at $E=100$ GeV
and $\sqrt{S}=1$ TeV for various angular cuts.
The percentual differences between the exact results and the two approximate
ones are also shown. Statistical errors are smaller that the last digit.}
\end{center}
\end{table}

Let us now consider the case when no angular cut is imposed on the scattered
electron, and the angular integration is performed over the whole phase space.
In this case, the minimum value of the photon virtuality $q^2_{min}$ can be
computed in a simple way by imposing that the invariant mass of the produced
hadronic system be bounded from below:
\beq
\label{q2min}
(q+k)^2\ge S_{min},
\eeq
where, for example, $S_{min}=4M^2$ in the case of heavy-quark pair production.
Equation~(\ref{q2min}) gives
\beq
q^2_{min}=-2k\cdot q + 4M^2 = -2k\cdot p y+4M^2\simeq -(ys-4M^2).
\eeq
The integration limits become
\beq
\label{intlim}
-\left(ys-4M^2\right)\leq q^2\leq -\frac{m_e^2y^2}{1-y}.
\eeq

In this case, however, we cannot simply take $q^2=0$ in $W_1(q^2,k\cdot q)$, as
we did in the previous case. In fact, processes involving the absorption of a
virtual photon by a hadron target are characterized by a dimensional parameter
$\lambda$, defined in such a way that the virtual photon cross section is
nearly equal to the photoproduction cross section for $\abs{q^2}<\lambda^2$,
while it falls rapidly to zero for virtualities above this value. In
particular, for the production of heavy objects, one can assume $\lambda=\xi
M$, where $\xi$ is a dimensionless parameter of order 1, since the
characteristic scale of the process is given by the mass of the heavy object.
For this reason, with quite a drastic approximation, instead of
eq.~(\ref{sigmagammap}) we may assume
\beq
\frac{W_1(q^2,k\cdot q)}{4k\cdot q}\simeq\sigma_{\gamma p}(q,k)
\Theta\left(q^2+\xi^2 M^2\right),
\eeq
which leads to
\beq
\label{wweq}
f_\gamma^{(e)}(y)=\frac{\aem}{2\pi}
\left[
2m_e^2y\left(\frac{1}{Q^2_{\sss WW}}-\frac{1-y}{m_e^2y^2}\right)
+\frac{1+(1-y)^2}{y}
\log\frac{Q^2_{\sss WW}(1-y)}{m_e^2 y^2}\right],
\eeq
where the quantity
\beq
Q^2_{\sss WW}\equiv {\rm min}(ys-4M^2,\xi^2 M^2)
\eeq
is usually referred to as the effective Weizs\"acker-Williams scale.

In practice, $ys-4M^2$ will also be of order $M^2$, because of the competing
effects of the large-$x$ suppression of the gluon density of the proton (which
favours the threshold region) and dynamical threshold suppression. We therefore
expect that in general the approximation
\beq
\label{appr}
Q^2_{\sss WW}=\xi^2 M^2
\eeq
will be appropriate. We see that, at least for the production of heavy objects,
the effective Weizs\"acker-Williams scale is determined by dynamical rather
than kinematical considerations. In the presence of an angular cut, dynamical
arguments are ineffective, because for small angles (which are of interest in
photoproduction processes) the minimum value of the virtuality,
eq.~(\ref{virtmin}), is well below $\xi^2 M^2$ in absolute value.

The parameter $\xi$ is essentially unconstrained (except that we expect it to
be of order 1). One can therefore absorb the non-logarithmic term in
eq.~(\ref{wweq}) by an appropriate redefinition of $\xi$. Consider for example
eq.~(\ref{wweq}) with only the most singular terms in $y$ taken into account.
We can get rid of the non-logarithmic term by replacing $\xi$ with $\xi'$,
where $\xi'$ is determined by the condition
\beq
-\frac{2}{y}+\frac{2}{y}\log\frac{\xi^2 M^2}{m_e^2 y^2}
=\frac{2}{y}\log\frac{{\xi'}^2 M^2}{m_e^2 y^2},
\eeq
or
\beq
\label{rhotoxi}
\xi'\simeq 0.6\,\xi.
\eeq

The numerical results for total cross sections at HERA, both for charm and
bottom production, are presented in table~\ref{tre}. In the first row we show
the result of the exact leading-order calculation. We see that the use of the
usual Weizs\"acker-Williams approximation of eq.~(\ref{WWequation}), shown in
the second row, leads to a completely wrong result. The following two entries
are obtained using eq.~(\ref{wweq}) with $Q^2_{\sss WW}=ys-4M^2$. The effect of
the inclusion of the non-logarithmic term is apparent, but still the
approximation is not satisfactory. The remaining entries are obtained by
implementing the dynamical considerations made above. We performed the
calculation using only the logarithmic term in the Weizs\"acker-Williams
distribution, and choosing different values for $\xi$, namely $\xi=1/2,1,2$.
The choice $\xi=1$ gives a very good agreement between the exact and the
approximated result. We then tested eq.~(\ref{rhotoxi}) by including the
non-logarithmic term in the Weizs\"acker-Williams function and taking $Q_{\sss
WW}=M/0.6$. Finally, the comparison between the entries in the fifth and the
last rows is a test of eq.~(\ref{appr}).

\begin{table}
\begin{center}
\begin{tabular}{|l||c|c|c|c|} \hline
& \multicolumn{2}{c|}{Charm}
& \multicolumn{2}{c|}{Bottom}
\\ \hline
    $Q_{\sss WW}$ & $\sigma$ (nb) & (\%) & $\sigma$ (nb) & (\%)
\\ \hline \hline
Exact result                & 353.4 & - & 3.135 & -
  \\ \hline
$\sqrt{yS-4M^2}$ only log   & 478.4 & 35.37 & 4.090 & 30.46
  \\ \hline
$\sqrt{ys-4M^2}$ only log   & 383.1 &  8.40 & 3.392 &  8.20
  \\ \hline
$\sqrt{ys-4M^2}$            & 367.9 &  4.10 & 3.265 &  4.15
  \\ \hline
$M$ only log                & 349.9 &  0.99 & 3.137 &  0.06
  \\ \hline
$M/2$ only log              & 327.1 &  7.44 & 2.934 &  6.41
  \\ \hline
$2M$  only log              & 372.6 &  5.43 & 3.339 &  6.51
  \\ \hline
$M/0.6$                     & 351.4 &  0.57 & 3.158 &  0.73
  \\ \hline
$min(\sqrt{ys-4M^2},M)$ only log & 349.0 & 1.24 & 3.128 & 0.22
  \\ \hline
\end{tabular}
\caption[]{\label{tre}
Total cross sections for heavy-flavour production at HERA, for different
choices of the Weizs\"acker-Williams distribution. The percentual differences
between the exact results and the approximate ones are also shown.}

\end{center}
\end{table}

\newpage
\begin{reflist}
\item\label{WWpaper}
	C.F. Weizs\"acker, \zp{88}{34}{612}; E.J. Williams, \pr{45}{34}{729}.
\item\label{HeraWorkshop}
   {\it Physics at HERA}, Proceedings of the Workshop, DESY, Hamburg,\\
    eds. W.~Buchm\"uller and G.~Ingelman (1991);\\
    A.C. Bawa and W.J. Stirling, {\it J.Phys.}{\bf G15}(1989)1339.
\item\label{CataniHautmann}
  	V.M.~Budnev et al., \prep{C15}{74}{181}; \\
        S.~Catani, M.~Ciafaloni and F.~Hautmann, \np{B366}{91}{135}.
\item\label{Laenen}
	E.~Laenen, S.~Riemersma, J.~Smith and W.L.~van Neerven,
	{\it Nucl. Phys.} {\bf B392} (1993) 162 and {\bf B392}(1993)229.
\item\label{MRSD0}
	A.~Martin, R.~Roberts and J.~Stirling, \pr{D47}{92}{867}.
\end{reflist}
\end{document}